\documentclass{JINST}
\usepackage{multirow}

\title{Irradiation of DEPFET-like transistors with Co-60 gamma source up to 10 Mrad}

\author{Pablo V\'azquez Regueiro\thanks{Corresponding author.}~, Eliseo P\'erez Trigo and Pablo Rodr\'iguez P\'erez\\
\llap IGFAE-University of Santiago de Compostela,\\
  Dto. F\'isica de Part\'iculas, Fac. F\'isica, Campus Vida s/n. 15782, Santiago de Compostela,  Spain\\
  E-mail: \email{pablo.vazquez@usc.es}}

\abstract{The Pixel Detector (PXD) of the Belle II experiment at superKEKB accelerator in Japan is based in the DEPFET technology. Two layers of 8+12 modules at a radius of 13 and 22 mm will give a spatial resolution below 10 $\mu$m. The radiation level expected in the first layer in ten years of operation is about 10 Mrad(Si) of total ionizing dose. In order to study the tolerance of the DEPFET technology sixty devices were irradiated using a standard procedure like $^{60}$Co gamma source. Different doping types, channel sizes and biasing conditions were studied.}


\begin{document}

\section{Introduction}

Depleted P- Channel Field Effect Transistor (DEPFET) active pixel detectors combine a first amplification stage with a fully depleted sensor in one single device, resulting in a very good signal-to-noise ratio even for thin sensors \cite{depfet}. DEPFET pixels are produced in MOS technology with two metal and two poly-silicon layers and have been developed for the use in X ray imaging and tracking in particle physics experiments \cite{laci} \cite{pablo}. BelleII experiment will set a DEPFET pixel layer at a distance of 13 mm from the interaction point integrating a total ionizing dose (TID) about 10 Mrad after 10 years of operation.
      
\begin{figure}[h]
\centering\begin{tabular}{c cl}
\includegraphics[width=.32\textwidth]{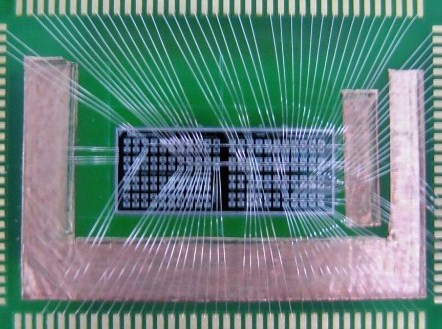} &
\includegraphics[width=.62\textwidth]{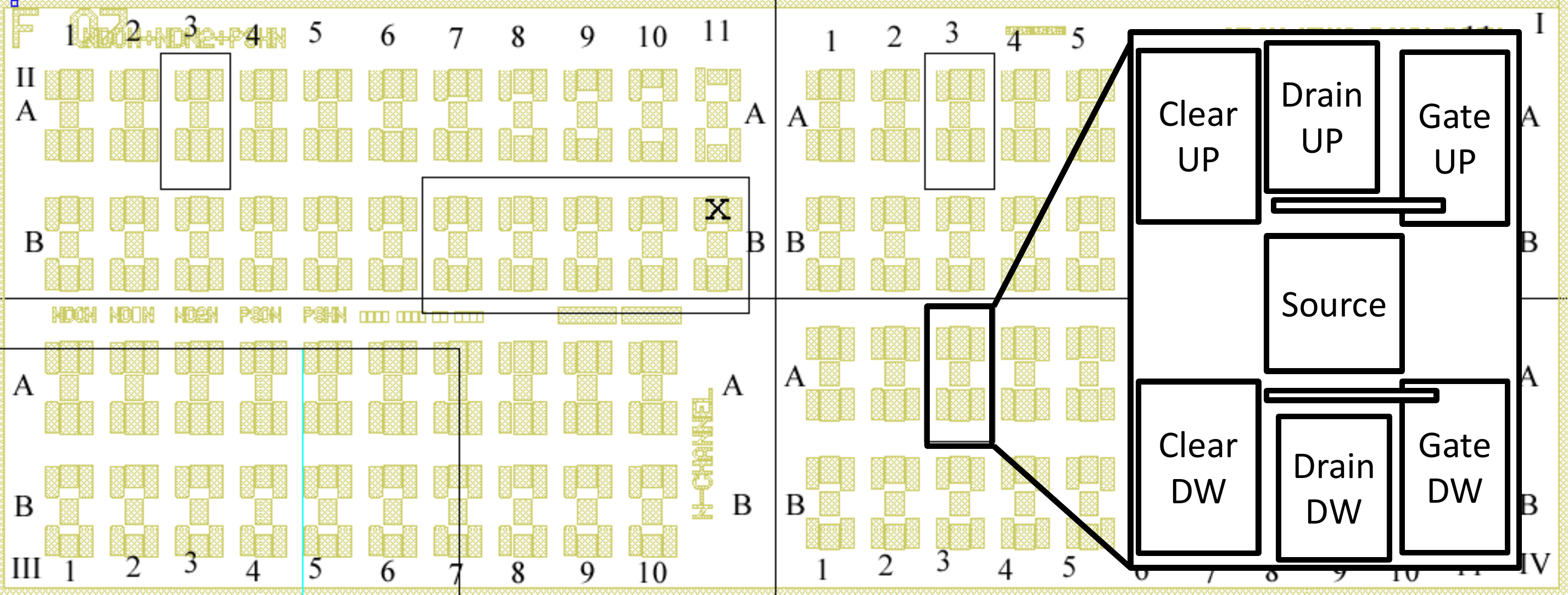} \\
\end{tabular}
\caption{Bonding detail picture and chip transistor layout.}
\label{figbon}
\end{figure}

Irradiation with ionizing and non-ionizing radiation has been performed on DEPFET structures up to 1 Mrad showing a threshold voltage shift about 5V. In order to characterize the PXD5 DEPFET technology \cite{testbeam} some chips with testing structures were designed. Two of those chips were used for radiation damage studies in Santiago. Two-transistor structures were grouped in quadrants, rows and columns, according to figure \ref{figbon}. Each quadrant had different doping profile while columns correspond to channel widths. Due to space constrains only 14, 22 and 24 transistors were instrumented from quadrants I, II and IV respectively with initial threshold voltages 3.5V, 0.5V and -2V. Threshold voltage shift, gain variation and sub-threshold region were studied with different biasing conditions. The influence of manufacturing and operational parameters such as doping, channel width and biasing voltage were studied. The irradiation took place in 11 steps (0.1 - 0.3 - 0.5 - 0.7 - 1 - 2 - 3 - 4 - 5.5 - 7.5 - 10 Mrad) plus an annealing of 28 days at room temperature during the summer 2010 in the Radiation Physics Laboratory of the Santiago de Compostela University. A radioactive source of 2080 Ci of activity was used allowing a dose on silicon of 11 krad/h.

\section{Measurements}

\begin{figure}[h]
\centering\begin{tabular}{c c}
\includegraphics[width=.47\textwidth]{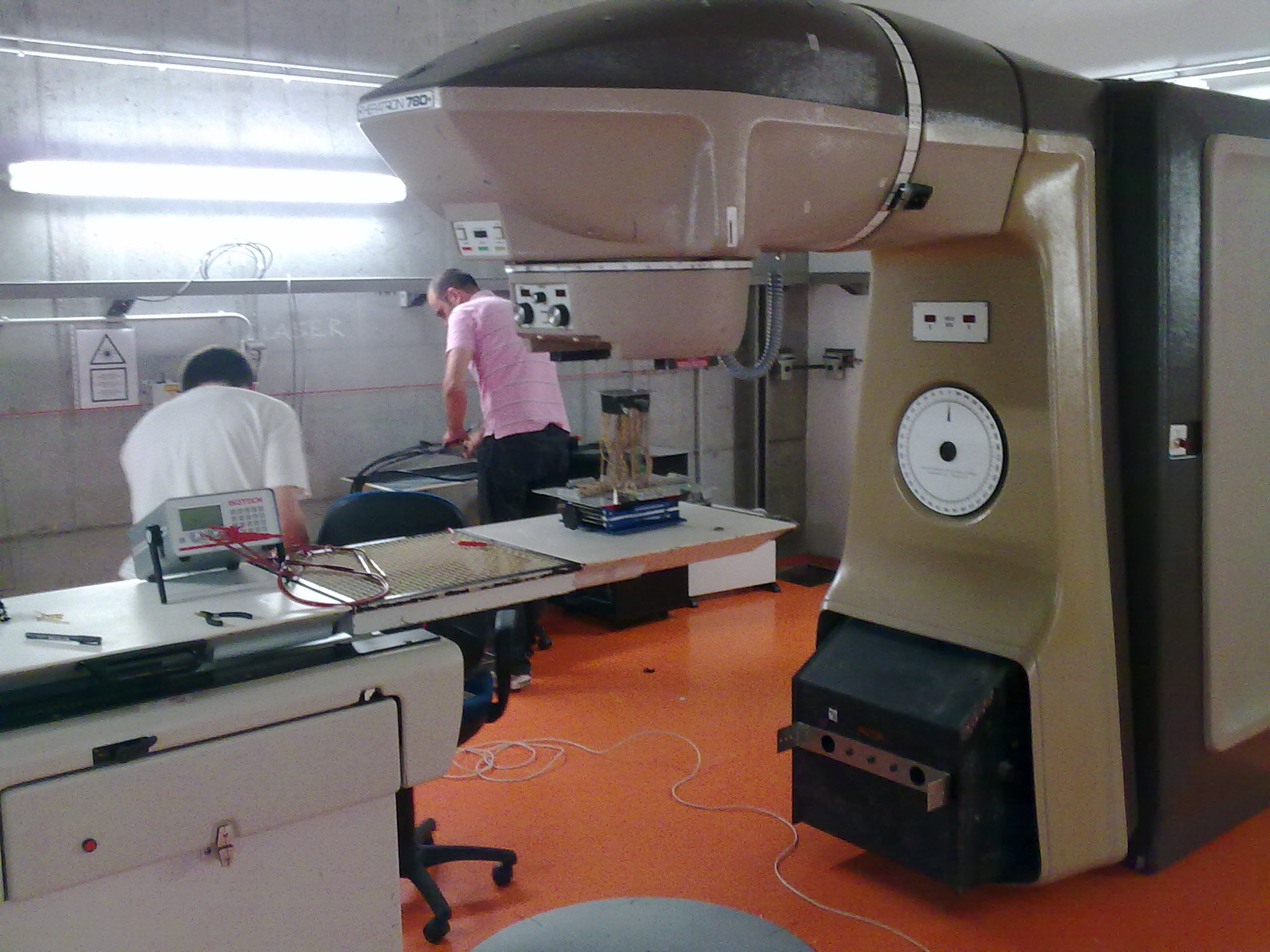} &
\includegraphics[width=.47\textwidth]{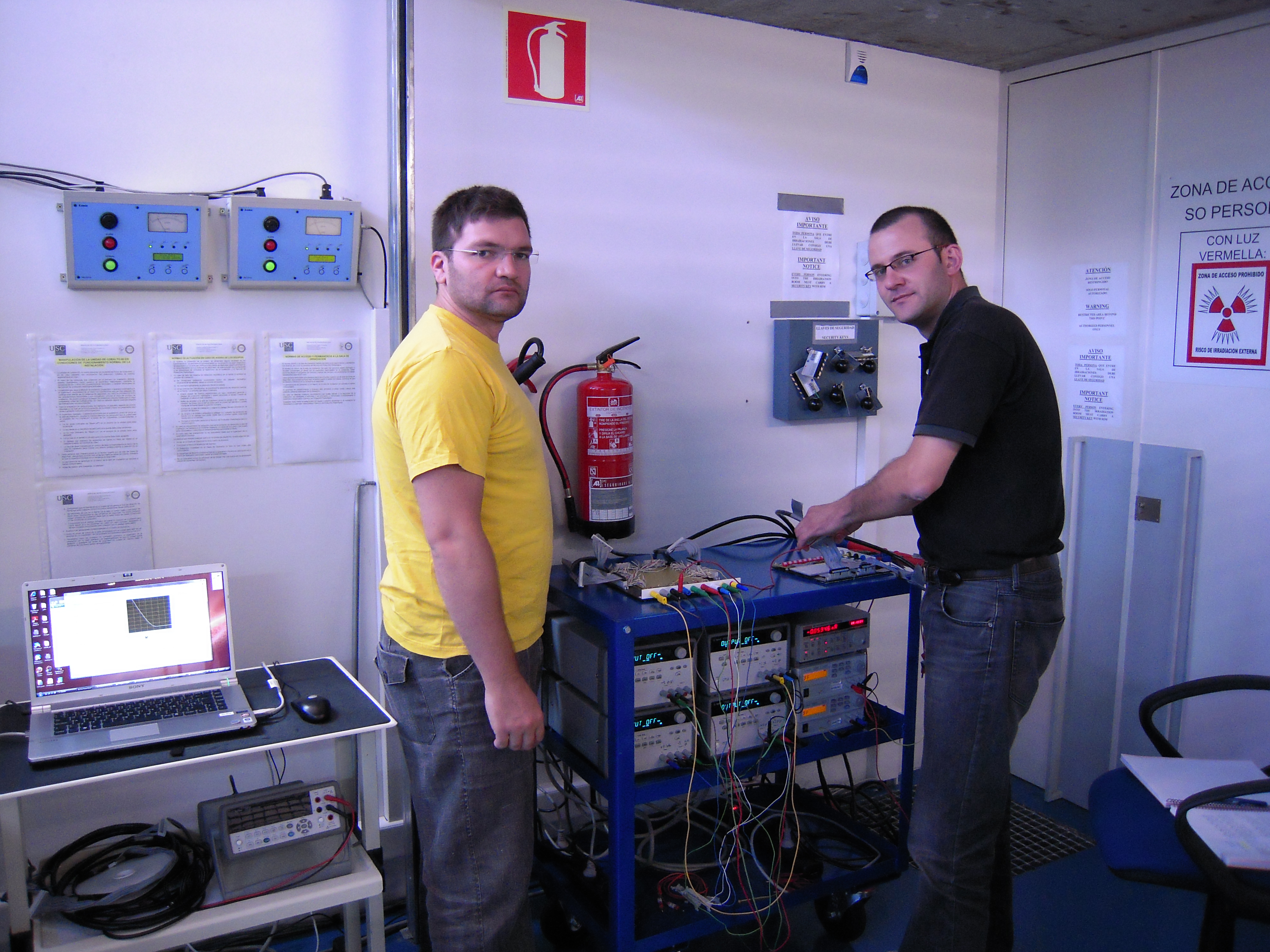}\\
\end{tabular}
\caption{$^{60}$Co source (inside the bunker) and biasing-testing setup (outside the bunker).}
\label{figset}
\end{figure}

The irradiation setup can be seen in figure \ref{figset}, the biasing and testing cards were placed outside the bunker. Ten meters in length cables allowed remote control. Electrostatic discharges caused during measurements made that only 22 transistors were operational at the end of the campaign, see table \ref{tabnam}.

\begin{figure}[h]
\centering\begin{tabular}{r l}
\includegraphics[width=.47\textwidth]{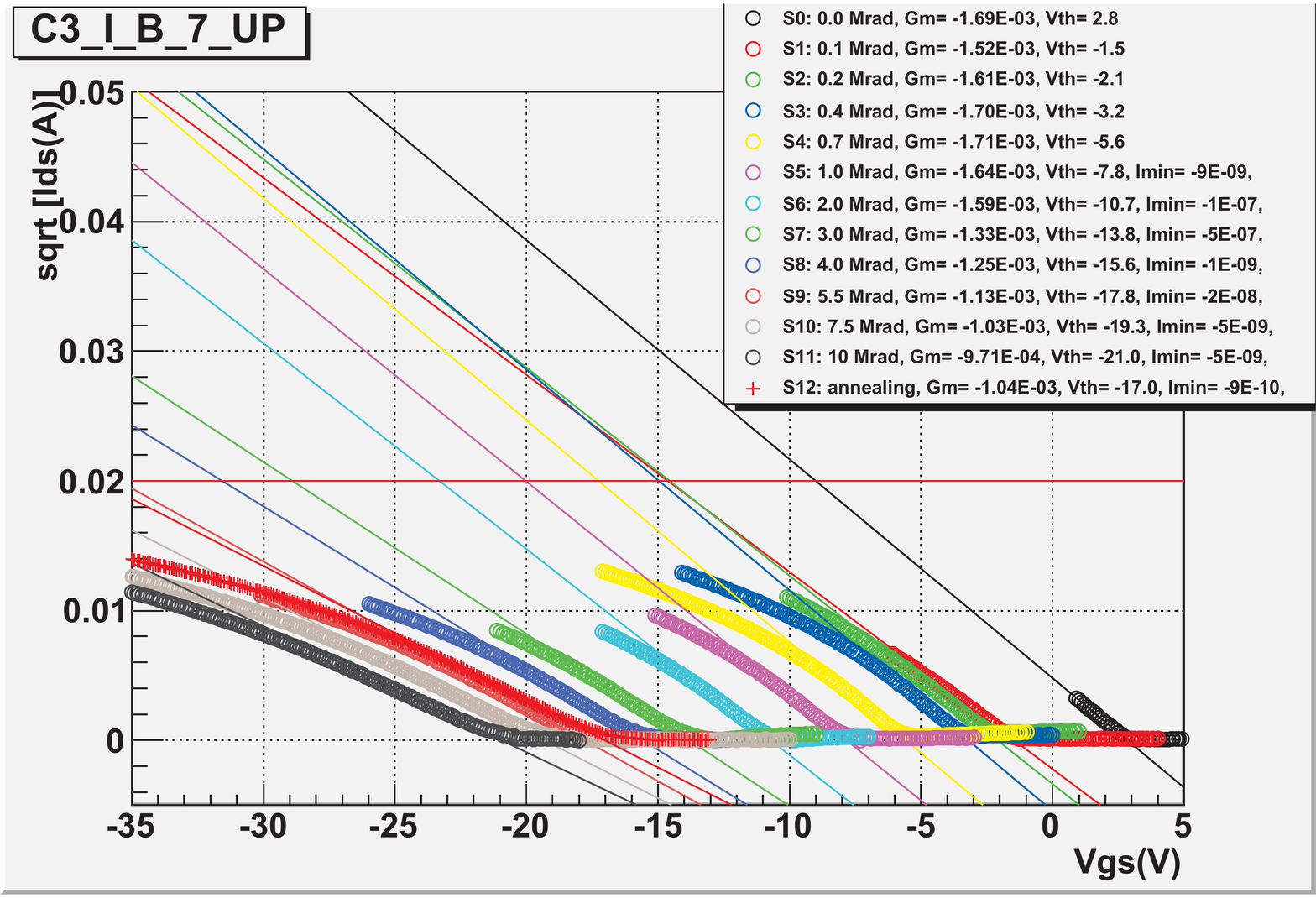} &
\includegraphics[width=.47\textwidth]{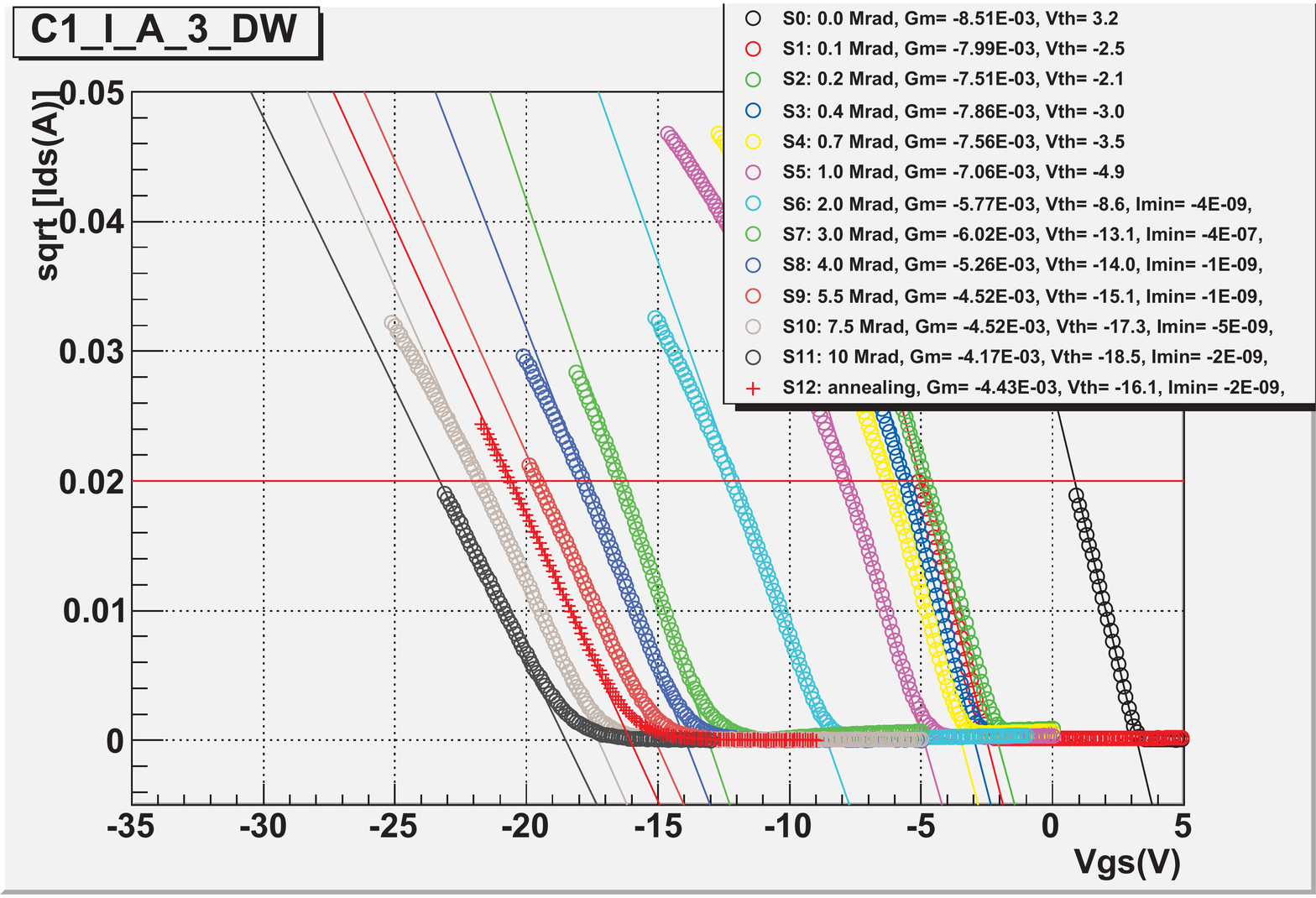}\\
\end{tabular}
\caption{Variation of $\sqrt{I_{ds}}/V_{gs}$ curve with dose for two transistors of gate widths 6 and 120 $\mu$m.}
\label{figIV}
\end{figure}

During irradiation all sources were grounded, bulk and clears were set to 10 V. Gate-source voltages were set according to table \ref{tabnam}. Some channels were in \emph{On state} before irradiation but after some dose all of them went to \emph{Off state} because $V_{gs}$ were not adapted to the threshold shift.
 The characterization procedure involved an IV curve measurement with a voltage step of 0.1 V, with a Keithley 487 picoammeter, in the following three hours of every step, except for steps 1 and 2 where the delay was 8 and 60 hours with the corresponding annealing. Temperature and relative humidity were monitored showing values in the range of $23^o$C and 46 \%rH. Threshold voltage, gain and sub-threshold region were calculated with the help of conduction and reverse linear fits of $\sqrt{I}V$ curves of figure \ref{figIV}. In the figures \ref{figVth}, \ref{figGain} and \ref{figsub} the annealing step was plotted in12 Mrad position.

\subsection{Threshold voltage}

\begin{figure}[h]
\centering\includegraphics[width=1\textwidth]{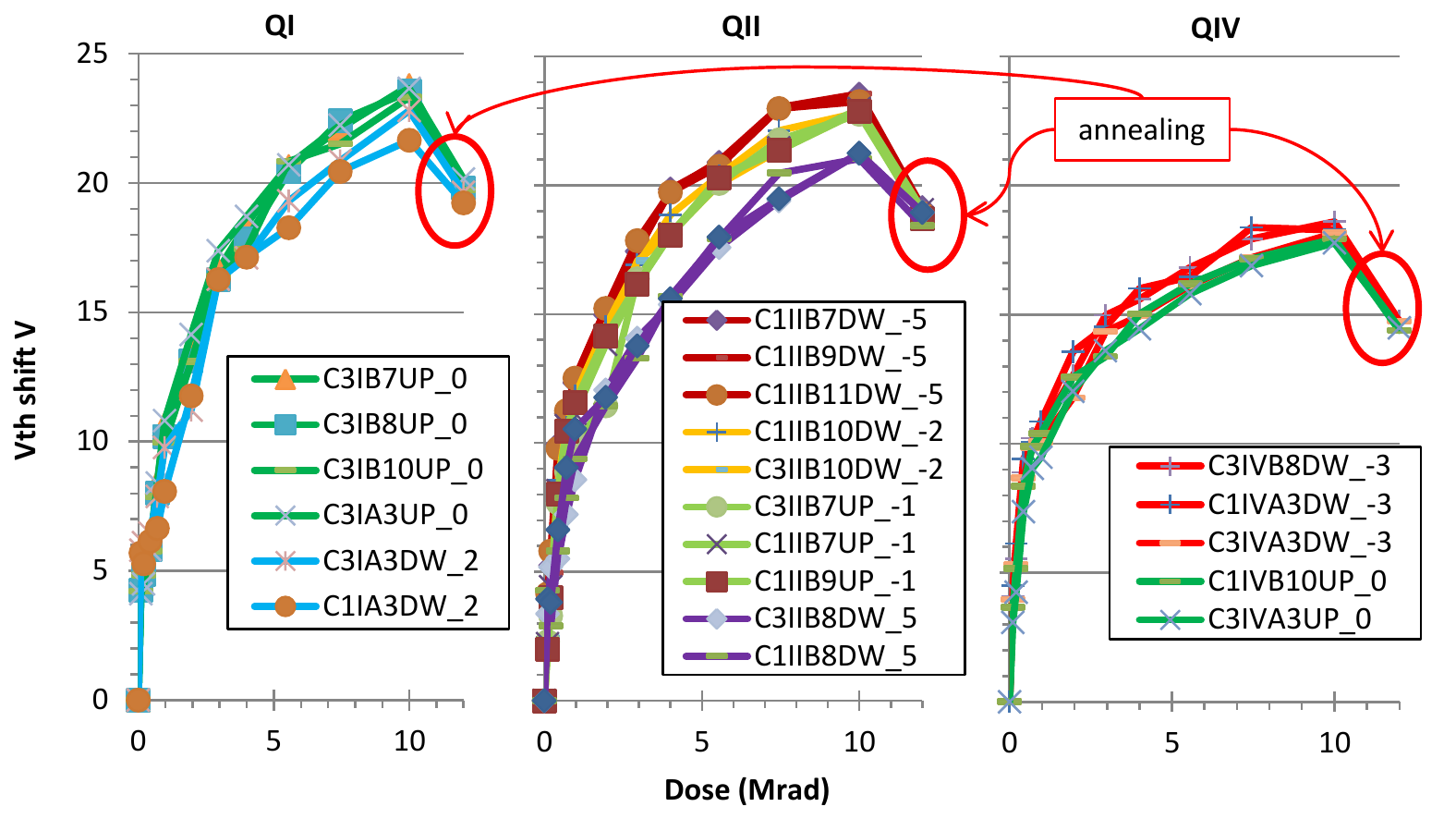}
\caption{Threshold voltage shift versus dose grouped by doping types (quadrants). Curves are color coded based on $V_{gs}$ : [0, 2] = [green,  blue] for QI. [-5, -2, -1, 5] = [red, orange, green , violet] for QII. [-3, 0] = [red, green] for QIV.}
\label{figVth}
\end{figure}

\begin{figure}[h]
\centering\includegraphics[width=.8\textwidth]{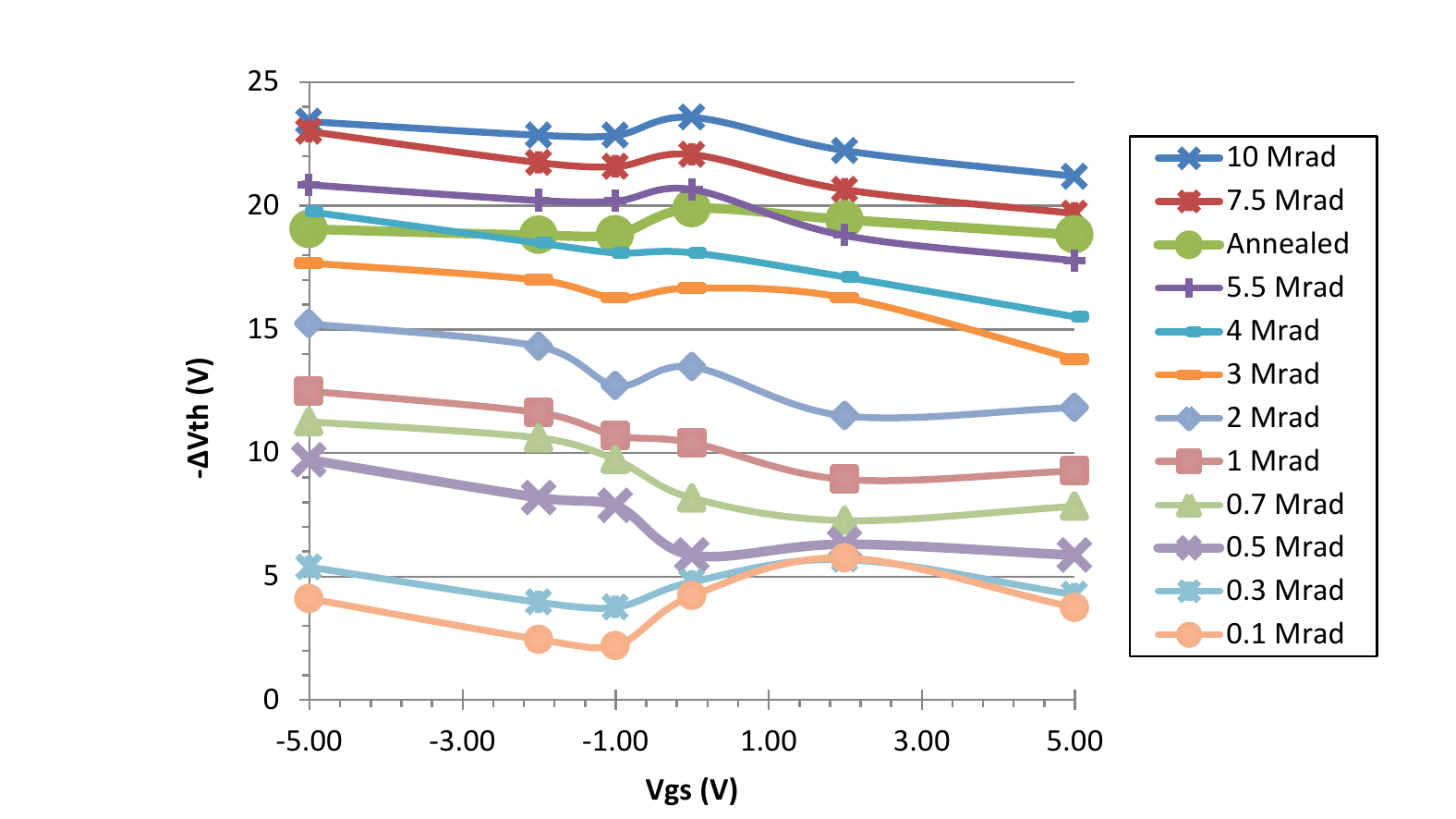}
\caption{Threshold voltage shift versus gate-source voltage.}
\label{figVgs}
\end{figure}

The threshold voltage is calculated as the intersection of the conduction and reverse regions linear fits in $\sqrt{I}V$ curves. Figure \ref{figVth} shows the threshold voltage shift versus the dose. The shift showed dependence with the doping profile and the gate-source voltage. Quadrants I and II having similar doping profile showed similar shift. After 10 Mrad it varied from -23 V for $V_{gs}$= -5 V to -21 V for $V_{gs}$= 5 V. Shifts were reduced to -19 V after the annealing. For quadrant IV, without internal gate implant, the shift is lower than quadrants I and II -18 V before and -14V after the annealing. The spread for similar transistors, with similar doping and $V_{gs}$, is below 0.5 V in all cases. Figure \ref{figVgs} details the threshold voltage shift dependence with the gate-source voltage. Only transistors with internal gate implant (QI, QII) were considered.

\subsection{Gain}

\begin{figure}[h]
\centering\includegraphics[width=1\textwidth]{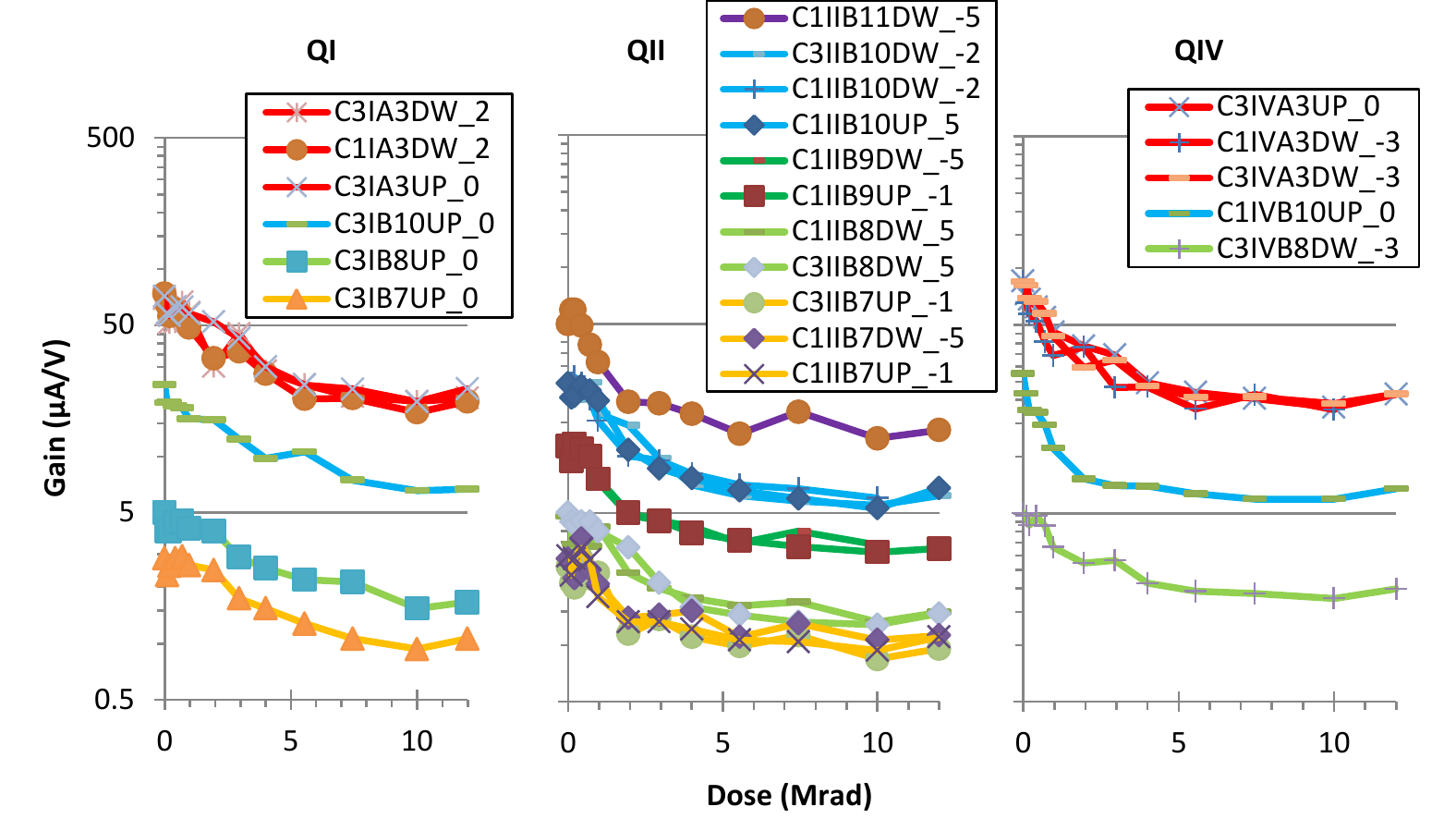}
\caption{Gain versus dose grouped by doping types. Curves are color coded based on channel width: [120, 40, 10, 6]=[red, blue, green, orange] for QI. [80, 40, 20, 10, 6] = [purple, blue, green, light green, orange] for QII. [120, 40, 10] = [red, blue, green] for QIV.}
\label{figGain}
\end{figure}

The gais is calculated as the slope of the conduction region linear fit in $\sqrt{I}V$ curves.
Figure \ref{figGain} shows the gain versus the dose. The gain is reduced on average by a factor of 3.7 after 10 Mrad. The reduction depends on the channel width ranging from 3 for w = 6 $\mu$m up to 4 for w > 40 $\mu$m. After the annealing the average gain reduction slightly recovers up to 3.4.

\subsection{Sub-threshold region}

\begin{figure}[h]
\centering\includegraphics[width=.75\textwidth]{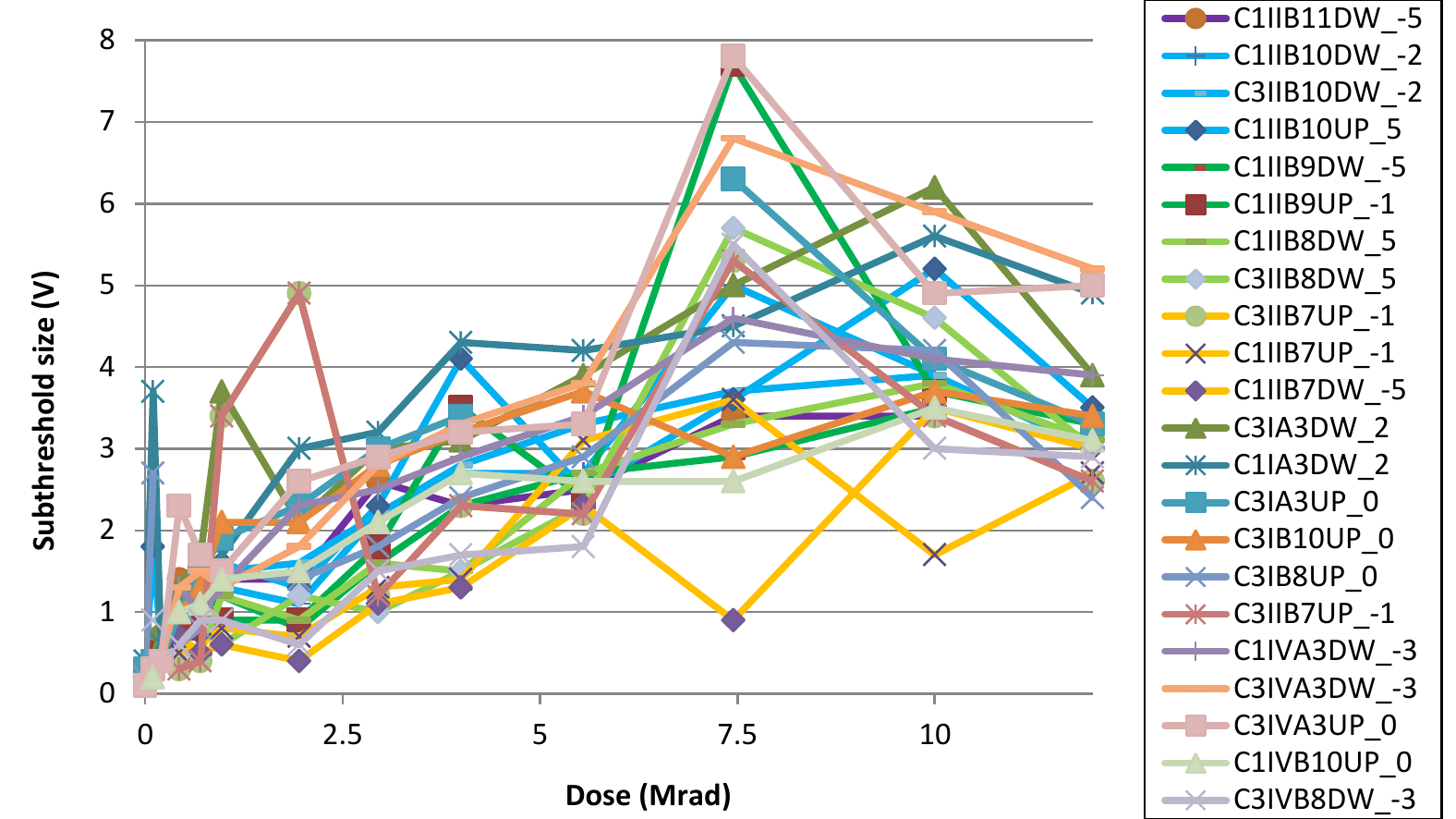}
\caption{Sub-threshold size versus dose for all transistors.}
\label{figsub}
\end{figure}

The method to calculate the sub-threshold region was to count  the number of points in the $\sqrt{I}V$ curve which do not correspond to a channel on or off behavior. The criteria was that the distance from the point to the conduction or reverse linear fits was bigger than 0.0001 in the $\sqrt{I}V$ space. Results are shown in figure  \ref{figsub}. The average value of sub-threshold region size for unirradiated transistors is 0.2 V, increased up to 4 V after 10 Mrad. As in the gain case, these values depend on the channel width, ranging from 2.5 V for the narrowest channel to 5 V of the widest one. The average size is reduced to 3.3 V after the annealing.

\section{Conclusions}

After being irradiated with a gamma source up to 10 Mrad, thick oxide DEPFET-like transistors have shown a negative threshold voltage shift of 18-23 V, reduced to 14-20 V after 28 days of annealing at room temperature. This voltage shift is foreseen to be compensated by changing the operation voltage in the switcher, the steering chip used for DEPFET matrix operation. In order to reduce this value, therefore the complexity of this chip, a new thin oxide technology is currently being developed in the DEPFET collaboration showing promising results in the first tests.

The gain is reduced by a factor of 3 - 4  and the sub-threshold size is increased by a factor of 10 - 30 after 10 Mrad depending on the channel width. The wider the channel the higher the effect. Both effects are relaxed after the annealing.

\begin{table}[h]
\centering\begin{tabular}{| c | c | l | c | c | c | c | c |}
\hline
Type & Vth$_0$ & Transistor &  W($\mu$m) &  L($\mu$m) & Vds & Vgs & State \\
\hline
\multirow{6}{*}{QI} & \multirow{6}{*}{3.5}
 & C2-I-B-7-UP & 6 & 6 & -5 & 0 & On $\rightarrow$ Off \\
 & & C2-I-B-8-UP & 10 & 6 & -5 & 0 & On $\rightarrow$ Off \\
 & & C2-I-B-10-UP & 40 & 6 & -5 & 0 & On $\rightarrow$ Off \\
 & & C1-I-A-3-DW & 120 & 6 & -5 & 2 & On $\rightarrow$ Off \\
 & & C2-I-A-3-UP & 120 & 6 & -5 & 2 & On $\rightarrow$ Off \\
 & & C2-I-A-3-DW & 120 & 6 & -5 & 2 & On $\rightarrow$ Off \\
\hline
\multirow{11}{*}{QII} & \multirow{11}{*}{0.5}
  & C1-II-B-7-UP & 6 & 6 & -5 & -1 & On $\rightarrow$ Off \\
 & & C1-II-B-7-WD & 6 & 6 & -5 & -5 & On $\rightarrow$ Off \\
 & & C2-II-B-7-UP & 6 & 6 & 0 & -1 & Off \\
 & & C1-II-B-8-DW & 10 & 6 &  -5 & 5 & Off \\
 & & C2-II-B-8-DW & 10 & 6  & 0 & 5 & Off \\
 & & C1-II-B-9-UP & 20 & 6   &-5 & -1 & On $\rightarrow$ Off \\
 & & C1-II-B-9-DW & 20 & 6   &-5 & -5 & On $\rightarrow$ Off \\
 & & C1-II-B-10-UP & 40 & 6  & -5 & -5  & On $\rightarrow$ Off \\
 & & C1-II-B-10-DW & 40 & 6  & -5 & -2  & On $\rightarrow$ Off \\
 & & C2-II-B-10-DW & 40 & 6 & 0 & -2  & Off \\
 & & C1-II-B-11-DW & 80 & 6 & -5 & -5 & On $\rightarrow$ Off \\
\hline
\multirow{5}{*}{QIV} & \multirow{5}{*}{-2}
& C2-IV-B-8-DW & 10 & 6 & -5 & -3 & On $\rightarrow$ Off \\
& & C1-IV-B-10-UP & 40 & 6 & 0 & 0 & Off \\
& & C1-IV-A-3-DW & 120 & 6 & 0 & -3 & Off \\
& & C2-IV-A-3-UP & 120 & 6 & -5 & 0 & Off \\
& & C2-IV-A-3-WD & 120 & 6 & -5 & -3 & On $\rightarrow$ Off \\
\hline
\end{tabular}
\caption{Irradiated transistors coded by \emph{chip}[C1,C2] - \emph{quadrant}[I,II,IV] - \emph{row}[A,B] - \emph{column}[3,7,8,9,10,11]- \emph{position}[UP,DW]}
\label{tabnam}
\end{table}

\acknowledgments

We want to thank Andreas Ritter and Rainer Richter from the MPI of Munich and Diego Miguel Gonz\'alez Casta\~no and Faustino G\'omez Rodr\'iguez from the Radiation Physics Laboratory of the Santiago de Compostela University for his support. This work was funded by the Spanish Ministry of Education and Science Project FPA2008-05979-C04-03.

\end{document}